\documentclass[aps,prc,amsmath,amssymb,onecolumn,showpacs,floatfix,superscriptaddress]{revtex4}
\usepackage{epsfig,graphics,color}
\usepackage{graphicx}
\usepackage{dcolumn}
\usepackage{bm}
\usepackage{amsmath}

\newcommand \tie {{\it i.e.}}

\newcommand \ra  {\rightarrow}

\newcommand \bvec{\left( \begin{array}{c} }
\newcommand \evec{\end{array} \right)}

\newcommand \eg {{\it e.g.}}
\newcommand \bea{\begin{eqnarray} }
\newcommand \eea{\end{eqnarray} }

\newcommand {\be} {\begin{equation}}
\newcommand {\ee} {\end{equation}}

\begin{document}
\title{Photon Production from Charge-Asymmetric Hot and Dense Matter}
\author{Guang-You Qin}
\affiliation{Physics Department, McGill University, 3600
University Street, Montreal, QC, H3A 2T8, CANADA}
\author{Abhijit Majumder}
\affiliation{Department of Physics, Duke University, Box 90305,
Durham, NC, 27708, USA}
\author{Charles Gale}
\affiliation{Physics Department, McGill University, 3600
University Street, Montreal, QC, H3A 2T8, CANADA}
\date{\today}
\begin{abstract}

A new channel of direct photon production from a quark gluon plasma (QGP) is explored. This process appears at Next-to-Leading-Order in the presence of a charge asymmetry in the heated matter and may be effectively described as the bremsstrahlung of a real photon from a thermal gluon. The photon production from this new mechanism is calculated in the effective theory of QCD at high temperature. The results show that the photon production rate may not as big as the annihilation and Compton scattering at low baryon density, but could become important in baryon-rich matter.

\end{abstract}


\maketitle
\section{INTRODUCTION}

It is the object of relativistic heavy-ion collisions to create and study strongly interacting matter excited beyond its hadronic phase~\cite{Harris:1996zx}. The existence of such a phase [the Quark Gluon Plasma (QGP)] has been predicted by Lattice QCD calculations~\cite{Karsch:2003jg} which exhibit a sudden rise in the scaled pressure and entropy density as the temperature is raised just beyond $T_c \sim 170$~MeV.  Detailed models of nuclear reactions had predicted that the energy deposition in the center-of-mass frame should be sufficient to cause temperatures at mid-rapidity in central collisions of gold nuclei to reach upwards of $300$~MeV~\cite{Wang:1996yf}. These predictions have been confirmed by the experimental results of the four Relativistic Heavy-Ion Collider (RHIC) detector collaborations, which have set a lower bound of about $5$~GeV/fm$^3$ on the energy density at a time $\tau = 1$~fm/c in central Au+Au collisions \cite{RHIC_Whitepapers}.

According to lattice calculations, such energy densities should place the excited matter firmly in the deconfined region. The fact that various lattice observables in the excited phase assume values close to those  expected for a free gas of quarks and gluons has let to the picture that beyond $T_c$, the degrees of freedom in chromodynamic matter are quasiparticles that carry the quantum numbers of  quarks and gluons. However, experimental results from the four RHIC detectors have cast doubts on this picture: the observed, large, elliptic and radial flow exhibited by the produced matter have led to the speculation that the produced matter may be strongly interacting~\cite{Gyulassy:2004zy}. These findings have given rise to phenomenological models in terms of bound states of quarks and gluons~\cite{Shuryak:2003ty,Shuryak:2004tx}. Such approaches, however, do not fare well in comparison with lattice susceptibility calculations~\cite{Koch:2005vg}. Such lattice comparisons do not reveal any information regarding the gluon sector of the plasma. As a result, efforts to elucidate the nature of the gluon structure have taken a phenomenological turn~\cite{Majumder:2007zh} \eg, there have been recent attempts to probe this structure through jet correlations~\cite{Koch:2005sx}. Here, we explore a novel means of probing the gluonic structure of the produced matter is proposed: through its possible electromagnetic signature.

Lepton pairs and real photons occupy a privileged status as they suffer essentially no final state interaction~\cite{Shuryak:1980tp} after their initial production. Thus, their emission rates have the potential to provide direct insight into the nature of the medium and its interactions. To this end, we focus on the electromagnetic signatures of  a series of pure glue processes, where the final rates are directly proportional to the gluon density of the produced matter.  Gluons do not carry electric charge, yet their interactions may generate electromagnetic signatures if the \emph{medium} is itself electrically charged. In partonic matter at equilibrium, this is achieved by the introduction of a non-vanishing charge chemical potential or a net asymmetry between the quark and anti-quark populations. This leads to a violation of Furry's theorem~\cite{Furry:1939qr} and the appearance of diagrams where two gluons may fuse through a quark loop to form a photon see Fig.~\ref{gg-gamma}. The possibility of such rates was first pointed out in Ref.~\cite{Majumder:2000jr}, the dilepton rates from such processes in different scenarios was expounded upon in Ref.~\cite{Majumder:2003vt}. Because of restrictions imposed by Yang's theorem~\cite{Yang:1950rg}, dilepton rates from such processes become appreciable only at high transverse momentum of the dilepton pair or in the case where the incoming gluons are themselves massive. The current work, in some ways an extension of these efforts will focus on the photon signature, which does not suffer from either of these constraints.

A large number of previous and even recent photon production calculations from an electrically charged QGP have neglected the baryon chemical potential $\mu_B$ (as well 
as other chemical potentials) in the plasma~\cite{Owens:1986mp, Kapusta:1991qp, Fries:2002kt,Arnold:2001ms}: in these cases, the photon production rate only depends on the temperature. It is now known that the central region at the CERN SPS and even RHIC is not just  heated vacuum~\cite{vonKeitz:1991kg, Olszewski:2002xk, Hammon:1999vw}, but actually displays a finite baryon density or an asymmetry between baryon and anti-baryon populations. Consequently, the baryon chemical density and thus $\mu_B$ in the QGP, does not vanish. In this case, the photon production rate (from the deconfined sector) is a function of both temperature $T$ and quark chemical potential $\mu$ of the QGP. 
In this treatment, isospin symmetry of two flavors is imposed, \tie, both $u$ and $d$ quarks will be assumed to be massless and have chemical potentials $\mu=\mu_B/3$. 
As can be immediately demonstrated, such a
plasma is globally electrically charged. The earliest estimates of photon production rates from electrically charged plasmas, in Ref.~\cite{Dumitru:1993us}, 
had pointed out that given an energy density $\epsilon$, 
the photon production rate will decrease strongly with increasing chemical potential of the medium. However, such calculations only include processes which are non-vanishing at $\mu_B=0$. The calculation of rates from these channels have been rigorously carried out in Ref.~\cite{Vija:1994is}, in the Hard-Thermal-Loop effective theory~\cite{Braaten:1989mz} at one loop. Rates at two loops in the HTL theory, at vanishing chemical potential,  were first presented in Ref.~\cite{Aurenche:1998nw}, where it was demonstrated that the two-loop rates from Bremsstrahlung processes actually dominate over rates at leading order in the coupling. All order resummed rates for Bremsstrahlung processes which includes the Landau-Pomeranchuck Migdal suppression from multiple scattering have been presented in Ref.~\cite{Arnold:2001ba}. The two loop rates have been  extended to finite chemical potential and chemical non-equilibrium in Ref.~\cite{Dutta:2001ii}. The effects of dynamical evolution of the quark gluon plasma on space-time integrated photon yields from such two loop  rates were presented recently in Ref.~\cite{Long:2005cn}.

It is the object of the current work to extend this line of inquiry and present rates for  hard photon production from processes which arise solely at finite chemical potential. As this is the first such attempt, we focus on establishing the basic processes and will pursue phenomenological applications elsewhere. In the following section, the calculation of the matrix element of the one-loop gluon-gluon-photon vertex in the HTL formalism is presented. In Sect.~III, the matrix element is incorporated into the hard photon production rate. In Sect.~IV, numerical estimates of photon rates from these processes are presented and compared with the leading order rates of Ref.~\cite{Kapusta:1991qp}. A comparison with the resummed rates of Ref.~\cite{Arnold:2001ba} cannot be performed as yet as these calculations have not been extended to finite baryon density. We summarize and outline future directions in Sect.~V.

\section{The One loop photon-gluon-gluon vertex}

\begin{figure}[htb]
\begin{center}\vspace{0cm}
\includegraphics[width=12cm]{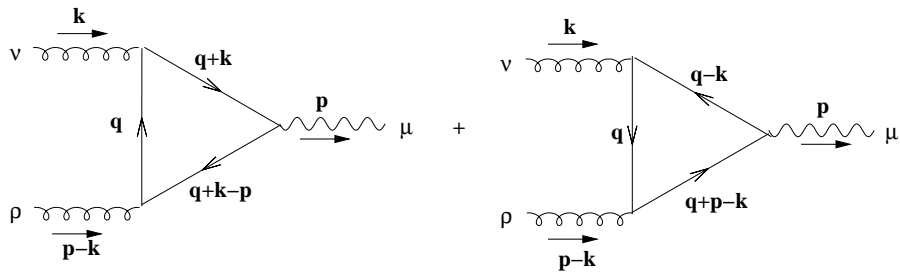}\vspace{0cm}
\end{center}
\caption{The one-loop Feynman diagrams of gluon-gluon-photon vertex as the sum of the two diagrams with quark numbers running in opposite directions in the quark triangle loops.} \label{gg-gamma}
\end{figure}

At zero temperature and at finite temperature and zero charge density, diagrams in QED that contain a fermion loop with an odd number of photon vertices (e.g. Fig.~\ref{gg-gamma}) are canceled by an equal and opposite contribution originating from the same diagram with fermion lines running in the opposite direction (Furry's theorem~\cite{Furry:1939qr,Itzykson01}). This statement can also be generalized almost unchanged to QCD, for processes with two gluons and an odd number of photon vertices.  A physical perspective is obtained by noting that all these diagrams are encountered in the perturbative evaluation of Green's functions with an odd  number of gauge field operators. At zero (finite) temperature, in the well defined  case of QED, the focus lies on quantities such as $\langle 0| A_{\mu_1} A_{\mu_2} ... A_{\mu_{2n+1}} |0\rangle $ ( $ Tr [ \rho(\mu,\beta)  A_{\mu_1} A_{\mu_2} ... A_{\mu_{2n+1}} ] $ ) under the action of the charge conjugation operator $C$. The photon, being charge conjugation negative, leads to $CA_{\mu}C^{-1} = -A_{\mu} $. In the case of the vacuum $|0\rangle $, we note that $C|0\rangle = |0\rangle$, as the vacuum is uncharged. As a result
\begin{eqnarray}
 \langle 0| A_{\mu_1} A_{\mu_2} ... A_{\mu_{2n+1}} |0\rangle
&=&  \langle 0| C^{-1}C A_{\mu_1} C^{-1}C A_{\mu_2} ... A_{\mu_{2n+1}} C^{-1} C |0\rangle
\nonumber \\
&=& \langle 0| A_{\mu_1} A_{\mu_2} ... A_{\mu_{2n+1}} |0\rangle (-1)^{2n+1}
= -\langle 0| A_{\mu_1} A_{\mu_2} ... A_{\mu_{2n+1}} |0\rangle = 0.
\end{eqnarray}
At a temperature $T$, the corresponding quantity to consider is
\begin{eqnarray}
\sum_{n} \langle n| A_{\mu_1} A_{\mu_2} ... A_{\mu_{2n+1}} |n\rangle
e^{-\beta (E_n - \mu Q_n)}, \nonumber
\end{eqnarray}
where $\beta = 1/T$ and $\mu$ is a chemical potential.
Here, however,
$C|n\rangle = e^{i\phi}|-n\rangle$, where $|-n\rangle$  is a state in the ensemble with the same number of antiparticles as there are particles in $|n\rangle$ and vice-versa. If $\mu = 0$ \tie, the ensemble average displays zero density, inserting the operator $C^{-1}C$ as before, one obtains,
\begin{eqnarray}
\langle n| A_{\mu_1} A_{\mu_2} ... A_{\mu_{2n+1}} |n\rangle
e^{-\beta E_n}
&=& - \langle -n| A_{\mu_1} A_{\mu_2} ... A_{\mu_{2n+1}} |-n\rangle
e^{-\beta E_n}.
\end{eqnarray}
The sum over all states will contain the mirror term $\langle -n| A_{\mu_1} A_{\mu_2} ... A_{\mu_{2n+1}} |-n\rangle e^{-\beta E_n} $, with the same thermal weight. As a result, summing over all states in the ensemble gives,
\begin{eqnarray}
\sum_{n} \langle n| A_{\mu_1} A_{\mu_2} ... A_{\mu_{2n+1}} |n\rangle
e^{-\beta E_n } = 0,
\end{eqnarray}
and Furry's theorem still holds. However, if $\mu \neq 0$ ($\Rightarrow$ unequal number of particles and antiparticles ) then
\begin{eqnarray}
\langle n| A_{\mu_1} A_{\mu_2} ... A_{\mu_{2n+1}} |n\rangle
e^{-\beta (E_n - \mu Q_n)}
= - \langle -n| A_{\mu_1} A_{\mu_2} ... A_{\mu_{2n+1}} |-n\rangle
e^{-\beta (E_n - \mu Q_n)},
\end{eqnarray}
the mirror term in such a case is $ \langle -n| A_{\mu_1} A_{\mu_2} ... A_{\mu_{2n+1}} |-n\rangle e^{-\beta (E_n + \mu Q_n)} $ , with a different thermal weight due to the fact that the net charge of the state $| -n \rangle$ is different and hence is weighted differently by the chemical potential. As a result, the thermal expectation of an odd number of gauge field operators is non-vanishing,
\begin{eqnarray}
\sum_{n} \langle n| A_{\mu_1} A_{\mu_2} ... A_{\mu_{2n+1}} |n\rangle
e^{-\beta (E_n - \mu Q_n)} \neq 0,
\end{eqnarray}
and Furry's theorem no longer holds. One may say that the medium, being charged, manifestly breaks charge conjugation invariance and these  Green's functions are thus finite, leading to the appearance of new processes in a perturbative expansion.  Two comments are in order: The eventual evaluation of the photon rate will appear to depend on the net baryon density 
(as opposed to the net quark density), as,  in the model of the plasma adopted, the net baryon density is carried equally by the up and down flavors. As pointed out in Ref.~\cite{Majumder:2003vt}, this is not the only means by which a finite charge density may be achieved. The processes of Fig.~\ref{gg-gamma} are also affected by the constraints imposed by Yang's theorem which states that a spin one particle may not decay or be formed by two identical massless vectors~\cite{Yang:1950rg}. The processes outlined in the following are computed within a thermalized medium, where the presence of longitudinal gluon excitations leads to a breaking of the symmetry which underlies Yang's theorem.

The incorporation of thermal gluon masses and self-energies in perturbation theory has to be done carefully, owing to issues arising from color gauge invariance. In this work, the calculation is carried out in the gauge invariant resummed theory of Hard-Thermal-Loops~\cite{Braaten:1989mz}, where one assumes the temperature $T \ra \infty$ and as a result the coupling constant $g(T) \ra 0$. Effective propagators and vertices involving soft $\sim gT$ momenta are obtained by integrating out the hard $\sim T$ modes. This allows for a well defined perturbative expansion of the photon production amplitude. The Feynman diagrams corresponding to the leading  contributions (in coupling) to the new channel of photon production are those of Fig.~\ref{gg-gamma}, with two gluons and a photon attached to a quark loop~\cite{Majumder:2000jr}.
Such a process does not exist at zero temperature, or even at finite temperature and vanishing chemical potential. At non-zero density, this leads to two new sources for photon production: The fusion of gluons to form a photon ($gg\to\gamma$) and the decay of a massive gluon into a photon and a softer gluon ($g\to g\gamma$).

The full, physical, matrix element is obtained by summing contributions from both diagrams which have fermion number running in opposite directions, $T^{\mu\nu\rho}(p,k,p-k) = T_1^{\mu\nu\rho}(p,k,p-k)+T_2^{\mu\rho\nu}(p,p-k,k)$. The amplitudes corresponding to these two diagrams may be expressed in the imaginary time formalism as:
\begin{eqnarray}
T_1^{\mu\nu\rho}(p,k,p-k) &=& T\sum_{q_0} \int {d^3q\over (2\pi)^3} eg^2 {\delta_{ab}\over 2} {\mathrm Tr}[\gamma^\mu\gamma^\alpha\gamma^\nu\gamma^\beta\gamma^\rho\gamma^\gamma] {(q+k)_\alpha q_\beta(q+k-p)_\gamma\over (q+k)^2 q^2(q+k-p)^2}\nonumber \\ T_2^{\mu\rho\nu}(p,p-k,k) &=& T\sum_{q_0} \int {d^3q\over (2\pi)^3} eg^2 {\delta_{ab}\over 2} {\mathrm Tr}[\gamma^\mu\gamma^\gamma\gamma^\rho\gamma^\beta\gamma^\nu\gamma^\alpha] {(q-k)_\alpha q_\beta(q-k+p)_\gamma\over (q-k)^2 q^2(q-k+p)^2}.
\end{eqnarray}
The masses of quarks have been omitted as the momenta of the quarks is considered hard $\sim T$ in the HTL expansion. In the imaginary time formalism, the zeroth components of four momentum are discrete Matsubara frequencies,
\begin{eqnarray}
q_0=i\omega_n+\mu=i(2n+1)\pi T+\mu\ ,\ k_0=i\omega_k=i2k\pi T\ ,\ p_0=i\omega_p=i2p\pi T,
\end{eqnarray}
where integers $n$, $k$ and $p$ in the above expression range from $-\infty$ to $\infty$, and $\mu$ is the quark chemical potential. It may be easily demonstrated~\cite{Majumder:2000jr}, using the the properties of the $\gamma$ matrices, that at zero chemical potential these two amplitudes cancel each other, consistent with the QCD generalization of Furry's theorem~\cite{Furry:1939qr,Itzykson01,Majumder:2000jr,Majumder:2003vt}. The sum over the Matsubara frequencies may be conveniently performed using the non-covariant propagator method of Ref.~\cite{Pisarski:1987wc,Majumder:2000jr}. Here, one defines a time-three-momentum propagator $\widetilde{\Delta}(\tau,E)$, as
\begin{eqnarray}
\widetilde{\Delta} ({{{i\omega }}_n}\pm\mu ,E) &=& {{{{\int }_0}}^{\beta }} d\tau {e^{{{{i\omega }}_n}\tau }}{{\widetilde{\Delta} }_{\pm}}(\tau ,E).
\end{eqnarray}
In the above equation, $E = |\vec{p}|$  represents the real energy of the particle from its three momentum and not the conjugate the imaginary time $\tau$. The explicit expression of the imaginary time quark propagator is given by 
\begin{eqnarray}
\widetilde{\Delta}_{\pm} (\tau,E) = \sum_{s=\pm 1}\widetilde{\Delta}_{s,\pm} (\tau,E) = \sum_{s=\pm 1} -\frac{s}{2E} \left[1-\widetilde{f}_\pm(sE)\right]{E^{-\tau ({sE}\mp\mu )}},
\end{eqnarray}
where $\widetilde{f}_\pm(E)= {1/(\exp[{{(E\mp\mu)/T}]+1})}$ are Fermi-Dirac distribution functions. Performing the summation of the Matsubara frequency $\omega_n$ leads to the expression for the amplitude:
\begin{eqnarray}
T^{\mu\nu\rho} &=& \int \frac{d^3q}{(2\pi )^3}{eg^2\delta_{ab}\over 2} {\mathrm Tr}[\gamma^\mu \gamma^\alpha\gamma^\nu\gamma^\beta\gamma^\rho\gamma^\gamma] \sum_{s_1s_2s_3} \frac{s_1s_2s_3}{8E_{q+k}E_qE_{q+k-p}} \nonumber \\ && {\tiny \frac{(q+k)_{s_1\alpha}{q_{s_2\beta }}{(q+k-p)_{s_3\gamma }}} {i\omega_p-s_1E_{q+k}+s_3E_{q+k-p}} \left(\frac{{\Delta\widetilde{n}}(E_{q+k})-{\Delta\widetilde{n}}({E_q})} {{{{i\omega}}_k}-{s_1E_{q+k}}+{s_2}{E_q}}- \frac{{\Delta\widetilde{n}}(E_{q+k-p})-{\Delta\widetilde{n}}({E_q})} {{i\omega_k-{{i\omega }}_p}-{s_3E_{q+k-p}}+{s_2}{E_q}}\right)}.
\end{eqnarray}
In the above equation, $q_s=(sE_q,\vec{q})$, and $\Delta\widetilde{n}(E)= \widetilde{f}_{+}(E)-\widetilde{f}_{-}(E)$.

The further evaluation of the photon production amplitude is carried out in the HTL approximation for the quark loop. In this limit, the photon and gluon momenta are considered soft $p,k \sim gT$, and the quark momenta are hard $q \sim T$, where $T$ is the temperature and $g$ is the effective coupling constant in the medium. The quark lines which carry a component of the external gluon energies are Taylor expanded as follows,
\begin{eqnarray}
E_{q+k} \approx
E_q+\vec{k}\cdot\hat{q}+{\vec{k}^2-(\vec{k}\cdot\hat{q})^2\over 2E_q}\ ,\ {(q+k)_{s\alpha}\over E_{q+k}} \approx \hat{q}_{s\alpha}+{k\over 2E_q}\hat{\mathcal{K}}_\alpha,
\end{eqnarray}
where $\hat{q}_s=(s,\hat{q})$ and $\hat{\mathcal{K}} = 2(0,\hat{k}-(\hat{k}\cdot\hat{q})\hat{q})$. Using the above approximation, allows for a factorization of the quark angular integral. Performing the integration over the magnitude of the quark momentum $\vec{q}$ analytically,  leads to the expression,
\begin{eqnarray}\label{amplitude}
T^{\mu\nu\rho} &=& \int {d\Omega_q\over (2\pi)^3} {eg^2\delta_{ab}\over 2} {\mu\over 8} {\mathrm Tr} [\gamma^\mu\gamma^\alpha\gamma^\nu\gamma^\beta\gamma^\rho\gamma^\gamma] \nonumber \\ && \left\{-{\hat{q}_{+\alpha}\hat{q}_{+\beta}\hat{q}_{+\gamma} \over p\cdot\hat{q}_+} \left({\vec{k}^2-(\vec{k}\cdot\hat{q})^2\over k\cdot\hat{q}_+} + {\vec{k}{'^2} -(\vec{k}'\cdot\hat{q})^2\over k'\cdot\hat{q}_+}\right)  - {k \hat{\mathcal{K}}_{\alpha}\hat{q}_{+\beta}\hat{q}_{+\gamma} - k'\hat{q}_{+\alpha}\hat{q}_{+\beta}\hat{\mathcal{K}'}_{\gamma} \over p\cdot\hat{q}_+}\left({\vec{k}\cdot\hat{q}\over k\cdot\hat{q}_+}-{\vec{k}'\cdot\hat{q}\over k'\cdot\hat{q}_+}\right) \right. \nonumber \\ && \left. -{\hat{q}_{+\alpha}\hat{q}_{+\beta}\hat{q}_{+\gamma} \over p\cdot\hat{q}_+} {[\vec{k}^2-(\vec{k}\cdot\hat{q})^2] - [\vec{k}{'^2}-(\vec{k}'\cdot\hat{q})^2]\over p\cdot\hat{q}_+}\left({\vec{k}\cdot\hat{q}\over k\cdot\hat{q}_+} -{\vec{k}'\cdot\hat{q}\over k'\cdot\hat{q}_+} \right) \right. \nonumber \\ && \left. - {\hat{q}_{+\alpha}\hat{q}_{+\beta}\hat{q}_{+\gamma} \over p\cdot\hat{q}_+} \left({\vec{k}\cdot\hat{q}\over k\cdot\hat{q}_+} {\vec{k}^2-(\vec{k}\cdot\hat{q})^2\over k\cdot\hat{q}_+} +{\vec{k}'\cdot\hat{q}\over k'\cdot\hat{q}_+} {\vec{k}{'^2}-(\vec{k}'\cdot\hat{q})^2\over k'\cdot\hat{q}_+} \right) + {2\hat{q}_{+\alpha}\hat{q}_{+\beta}\hat{q}_{+\gamma} \over p\cdot\hat{q}_+} \left({(\vec{k}\cdot\hat{q})^2\over k\cdot\hat{q}_+}+{(\vec{k}'\cdot\hat{q})^2\over k'\cdot\hat{q}_+} \right) \right. \nonumber \\ && \left. - \hat{q}_{-\alpha}\hat{q}_{+\beta}\hat{q}_{+\gamma} {\vec{k}'\cdot\hat{q}\over k'\cdot{q}_+} - \hat{q}_{+\alpha}\hat{q}_{-\beta}\hat{q}_{+\gamma} {\vec{p}\cdot\hat{q}\over p\cdot\hat{q}_+} - \hat{q}_{+\alpha}\hat{q}_{+\beta}\hat{q}_{-\gamma} {\vec{k}\cdot\hat{q}\over k\cdot\hat{q}_+}  \right\},
\end{eqnarray}
where $k'=p-k$ and $d\Omega_q=d\cos d\theta_q d\phi_q$ is the differential solid angle of the quark momentum $\vec{q}$. The first line of the above equation, demonstrates explicitly that the amplitude is directly proportional to the chemical potential $\mu$. As a result, the contribution to the photon production rate from this channel will grow quadratically with increasing chemical potential if the temperature of the medium is held constant.

The remnant angular integral over $d\Omega_q$ is nontrivial and is performed numerically. The possibility of radiation or absorption of a space-like gluon by an on-shell quark induces a long distance enhancement  in a small part of phase space in each of the two diagrams separately. Such contributions are diminished by the destructive interference between the two diagrams and thus do not contribute to any eventual long distance enhancement in the rate of photon production from this channel.  Including all contributions, leads to the survival of only the imaginary part of the amplitude in this sector. The resulting expression is,
\begin{eqnarray}
T^{\mu\nu\rho} &=& \int {d\phi\over (2\pi)^3} {eg^2\delta_{ab}\over 2} {\mathrm Tr} [\gamma^\mu\gamma^\alpha\gamma^\nu\gamma^\beta\gamma^\rho\gamma^\gamma] \left(-{\mu\omega\over 8k}\right){i\pi}  \nonumber\\ && \left\{ -{1\over p\cdot\hat{\bar{q}}} \left( k\hat{\mathcal{K}}_{\alpha} \hat{\bar{q}}_{+\beta} \hat{\bar{q}}_{+\gamma} - k'\hat{\bar{q}}_{+\alpha} \hat{\bar{q}}_{+\beta} \hat{\mathcal{K}'}_{\gamma} \right)-{[\vec{k}^2-(\vec{k}\cdot\hat{\bar{q}})^2] - [\vec{k}{'^2}-(\vec{k}'\cdot\hat{\bar{q}})^2]\over (p\cdot\hat{\bar{q}}_+)^2}\hat{\bar{q}}_{+\alpha}\hat{\bar{q}}_{+\beta} \hat{\bar{q}}_{+\gamma}\right.\nonumber\\ &&\left. + {k\over p\cdot\hat{\bar{q}}}\left(\hat{\mathcal{Q}}_{\alpha} \hat{\bar{q}}_{+\beta} \hat{\bar{q}}_{+\gamma} + \hat{\bar{q}}_{+\alpha} \hat{\mathcal{Q}}_{\beta} \hat{\bar{q}}_{+\gamma} + \hat{\bar{q}}_{+\alpha} \hat{\bar{q}}_{+\beta} \hat{\mathcal{Q}}_{\gamma}\right) - \hat{\bar{q}}_{+\alpha} \hat{\bar{q}}_{+\beta} \hat{\bar{q}}_{-\gamma} \right\},
\end{eqnarray}
where $\hat{\bar{q}}$ and $\mathcal{Q}$ are given by
\begin{eqnarray}
\hat{\bar{q}}_s &=& \left(s, \sqrt{1-{k_0^2/ \vec{k}^2}}\cos\phi, \sqrt{1-{k_0^2/ \vec{k}^2}}\sin\phi, {k_0/ k}\right) \nonumber \\ \hat{\mathcal{Q}} &=& \left(0, -{k_0/ k} \sqrt{1-{k_0^2/ \vec{k}^2}}\cos\phi, -{k_0/ k} \sqrt{1-{k_0^2/ \vec{k}^2}}\sin\phi, 1-{k_0^2/ \vec{k}^2}\right).
\end{eqnarray}
The above equations represent the central result of this effort. While this been derived in a thermalized environment, it may be easily generalized to moderate departures from equilibrium. This will remain the subject of a future effort, where the above matrix element will be used to study the photon signature emanating from the gluon sector of different models of the QGP. In the subsequent sections, the above amplitude will be applied to compute the photon production rate from gluon fusion and decay in the simplest model of a QGP, a plasma of quasi-particle quarks and gluons in complete thermal and chemical equilibrium.

\section{The photon self-energy and the photon production rate}

\begin{figure}[htb]
\begin{center}\vspace{0.0cm}
\includegraphics[width=5cm]{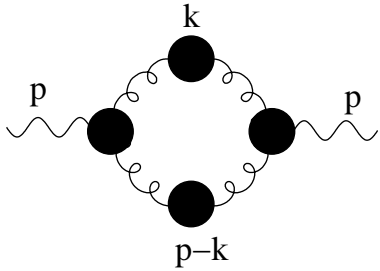}\vspace{0cm}
\end{center}
\caption{The Feynman diagram of the photon self-energy evaluated in the work.} \label{photon-SF}
\end{figure}

\begin{figure}[htb]
\begin{center}\vspace{0.0cm}
\includegraphics[width=8cm]{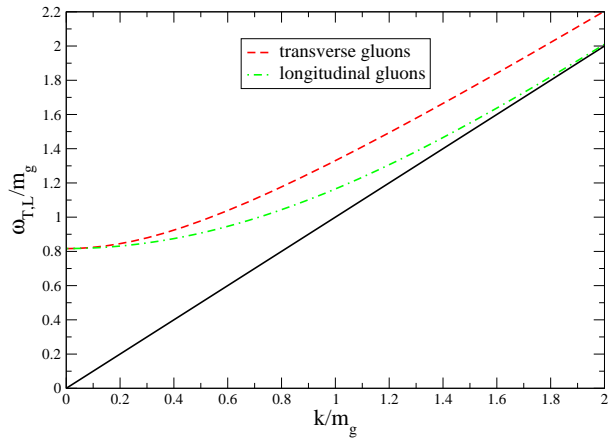}\vspace{0cm}
\end{center}
\caption{The dispersion relations $\omega_{T,L}(k)$ for transverse gluons and longitudinal gluons in a quark gluon plasma, where $m_g$ is the thermal gluon mass.} \label{dispersion-relation}
\end{figure}

In the preceding sections, the origin and derivation of the amplitude of photon production from two gluons through a quark triangle has been outlined. Such an amplitude may be used for different processes such as the fusion of gluons to form a photon or the decay of a gluon into a photon and a gluon of lower energy. The production or absorption rate of photons from a dense medium is related to the imaginary part of the photon self-energy in the medium. The choice of self-energy depends on the process of interest. In what follows, the focus will lie on the imaginary part of the photon self-energy of Fig.~\ref{photon-SF}.
In the case of a medium in complete thermal and chemical equilibrium, the thermal photon emission rate $R=d^4N/d^4x$ is related to the discontinuity or the imaginary part of the retarded photon self-energy $\Pi^R_{\mu\nu}$ at finite temperature $T$ through the relation~\cite{KG,McLerran:1984ay, Gale:1990pn},
\begin{eqnarray}
E{dR\over d^3p} = -{1 \over (2\pi)^3} \mathrm{Im} \Pi_{\mu}^{R,\mu} {1\over e^{E\over T}-1},
\end{eqnarray}
where $E$ and $p$ are the energy and momentum of the photons. This formula is valid to all orders of strong interactions, but only to $e^2$ in the electromagnetic interactions, as the photons, once produced, will tend to escape from the matter without further interaction. The photon self energy from Fig.~\ref{photon-SF} may be expressed formally as,
\begin{eqnarray}
\Pi^{\mu\mu'}(p) = T\sum_{k_0}\int{d^3k\over (2\pi)^3}T^{\mu\nu\rho}(p,k,p-k)S_{\nu\nu'}(k) T'^{\mu'\nu'\rho'}(-p,-k,-p+k)S_{\rho\rho'}(p-k),
\end{eqnarray}
where $T^{\mu\nu\rho}(p,k,p-k)$ is the effective photon-gluon-gluon vertex evaluated in the last section, and $S_{\mu\nu}(k)$ is the effective gluon propagator, after summing up all the HTL contributions to the self-energy of the gluon. In the Coulomb gauge, the propagator is given by~\cite{KG}
\begin{eqnarray}
S_{\mu\nu}(k) = {1\over F_T-k^2}P^T_{\mu\nu}(k) + {1\over F_L-k^2}{k^2 \over {\vec{k}^2}} u_{\mu}u_{\nu}
\end{eqnarray}
In the above equation, $P^T_{00}(k)=0$, $P^T_{ij}(k)=\delta_{ij}-k_ik_j/\vec{k}^2$ is the transverse projection tensor and $u^\mu=(1,0,0,0)$ specifies the rest frame of the medium; the explicit expressions for $F_T$ and $F_L$ can be found in Ref.~\cite{KG}. It is convenient to define the transverse and longitudinal gluon propagators $\Delta_T(k_0,k)={1/(F_T-k^2)}$ and $\Delta_L(k_0,k)={1/(F_L-k^2)}{k^2 \over {\vec{k}^2}}$. In the complex $k_0$ plane, these propagators exhibit a discontinuity or cut from $-k$ to $k$; in addition, they have poles at $k_0=\pm\omega_{T,L}(k)$, which give the two dispersion relations for the longitudinal and transverse modes of the gluons in the medium. In the interest of completeness, these are plotted in Fig.~\ref{dispersion-relation}.
In the plot, the upper branch is the dispersion relation for transverse excitation modes, and the lower branch is for the the longitudinal one. The solid line represents the light cone. The dispersion relations admit a thermal gluon mass at the intercept $k=0$ given as $m_g^2= C_A g^2 T^2/6+N_F(T^2+{3\mu^2/ \pi^2})/12$~\cite{KG}.

In order to calculate the thermal photon differential production rate, we evaluate the imaginary part or the discontinuity of the photon self-energy, which involves evaluating its various cuts. In the interest of a physical interpretation of the various cuts, the polarization tensor $P^T_{\mu\nu}(k)$ may be expanded as a product of polarization vectors as $ P^T_{\mu\nu}(k) = \epsilon_{+\mu}(k) \epsilon_{+\nu}^*(k) +\epsilon_{-\mu}(k) \epsilon_{-\nu}^*(k)$, where the $z$-axis is chosen as the direction of the photon momentum $\vec{k}$. Then the effective propagator may be formally written as
\begin{eqnarray}
S_{\mu\nu}(k) = \sum_{i=+,-,0} \Delta_i(k) \epsilon_{i\mu}(k)\epsilon_{i\nu}(k),
\end{eqnarray}
where we set $\epsilon_0^\mu = u^\mu = (1,0,0,0)$ and $\Delta_\pm=\Delta_T$, $\Delta_0=\Delta_L$. The entire expression for the rate, may be expressed in a factorized form $f(k_0)g(p_0-k_0)$, where $k^0$ is the Matsubara frequency of the gluon and $p^0$ is the frequency of the external photon. The remaining sum over $k^0$ and the discontinuity across the real $p^0$ is achieved by means of the identity,~\cite{Braaten:1990wp,Majumder:2000jr},
\begin{eqnarray}
\mathrm{Disc} T\sum_{k_0} f(k_0)g(p_0-k_0) &=& 2\pi i \int d\omega \int d\omega'  \rho_1(\omega)\rho_1(\omega') \delta(\omega+\omega'-E)(1+f(\omega)+f(\omega')),
\end{eqnarray}
where $f(\omega)$ and $f(\omega')$ are the distribution functions of gluons, and $\rho_1(\omega)$ and $\rho_2(\omega')$ are the spectral functions of $f(k_0)$ and $g(p_0-k_0)$. The spectral function $\rho(z)$ for $f(z)$ is defined as $\rho(z)=\mathrm{Disc}f(z)/(2\pi i)$. Employing the above formula, the discontinuity in the photon self-energy may be expressed in a kinetic form,
\begin{eqnarray}
E{dR\over d^3p} &=& \sum_{l=\pm}\sum_{i,j=\pm,0} {1\over 2(2\pi)^2} {1\over e^{E\over T}-1} \int{d^3k\over (2\pi)^3}  \int d\omega \int d\omega' \rho_i(\omega)\rho_j(\omega') \delta(\omega+\omega'-E) (1+f(\omega)+f(\omega')) |M_{lij}|^2,
\end{eqnarray}
where the matrix element $M_{lij}=\epsilon_{l\mu}(p)\epsilon_{i\nu}(k)\epsilon_{j\rho}(p-k)T^{\mu\nu\rho}(p,k,p-k)$, and $\rho_i(\omega)$ and $\rho_j(\omega')$ are the spectral functions of the gluon propagators $\Delta_i$ and $\Delta_j$.
\begin{eqnarray}
\rho_{T,L}(k_0,k) = Z_{T,L}(k)[\delta(k_0-\omega_{T,L}(k))-\delta(k_0+\omega_{T,L}(k))] +\beta_{T,L}(k_0,k)\theta(k^2-k_0^2).
\end{eqnarray}
The spectral functions $\rho$ contain contributions from the poles $\omega_{T,L}$ with residue $Z_{T,L}$ as well as from the branch  cuts $\beta_{T,L}$. The product of two $\rho$ functions give three types of contribution: pole-pole, pole-cut, and cut-cut. The pole-pole terms represent the process involving two quasi-particles with dispersion relations displayed in Fig.~\ref{dispersion-relation}. The terms from the cuts represents the processes involving space-like gluons from the medium, \tie., gluons which are intermediate states of a scattering process. In this first effort, the focus will lie on the hard photon production rate, \tie, photons with momenta $p \sim T$. This requires that at least one of the gluons in Fig.~\ref{photon-SF} to be hard. While in the usual HTL prescription, such particles receive suppressed contributions from hard loops, a component of the HTL self-energy which produces a thermal gluon mass is retained. The cut-cut contribution with two space-like gluons is dominant only in the region where both gluon momenta are soft and may be ignored in this effort. As a result, the two main contributions to the hard photon rate computed in this paper emanate form the pole-pole and pole-cut terms. In the subsequent section, numerical estimates of the photon rate from a set of production scenarios is provided and compared to the corresponding rate from the leading process of hard photon production from Compton scattering and pair annihilation.

\section{Results}

\begin{figure}[htb]
\begin{center}\vspace{0.0cm}
\includegraphics[width=10cm]{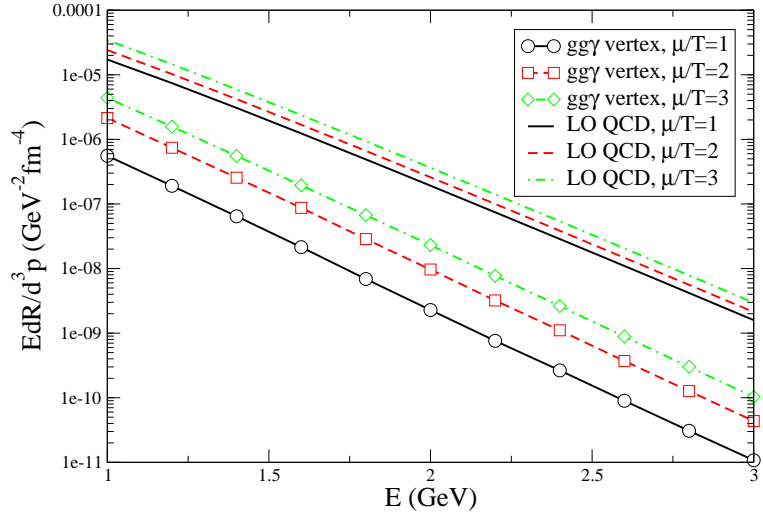}\vspace{0cm}
\end{center}
\caption{The differential rate of photons from $gg\gamma$ vertex in a hot and dense medium with temperature $T=200\mathrm{MeV}$ compared with the contribution from QCD annihilation and Compton processes.} \label{T=200}
\end{figure}

In this section, numerical results for the hard photon production rate from a plasma with a finite charge density will be presented. The calculation is performed for two massless flavors of  quarks with $\mu_u=\mu_d=\mu_B/3$. In such a plasma, the strong coupling constant $\alpha_s=0.4$. The strange sector has been ignored in this calculation. In Fig.~\ref{T=200}, the photon production from our new channel is compared with the contribution from the leading order QCD processes of quark anti-quark annihilation and quark gluon Compton scattering.
The photon differential rate from QCD annihilation and Compton processes at finite temperature and chemical potential is parameterized as in Ref.\cite{Kapusta:1991qp, Vija:1994is} by
\begin{eqnarray}
E{dR\over d^3p} = {5\over 9} {\alpha \alpha_s \over 2 \pi^2} \left(T^2+{\mu^2\over \pi^2}\right) e^{-{E\over T}} \ln\left({2.912ET\over g^2(T^2+{\mu^2/ \pi^2})}+G\right),
\end{eqnarray}
where the dimensionless quality $G$ is fitted to be $G=\ln(1+\mu^2/\pi^2T^2)$ for $\mu/T \leq 1$ and
$G=\ln(1+0.139\mu^2/T^2)$ for $\mu/T \geq 1$. One may immediately note from Fig.~\ref{T=200} that the contribution from the  new channels presented in this paper, the $gg\gamma$ vertex, to the photon production is much smaller than the QCD annihilation and Compton processes at low chemical potential. However,  with increasing chemical potential at a fixed temperature, the photon production rate from the $gg\gamma$ vertex tends to increase at a swifter rate than QCD annihilation and Compton contribution. This may be understood from the fact that the matrix element corresponding to the $gg\gamma$ vertex in Fig.~\ref{gg-gamma} is proportional to the chemical potential as we see from Eq.~\ref{amplitude}. This leads to the conclusion that in baryon-rich matter such as that produced in low energy collisions of heavy ions or in the core of neutron stars, where the chemical potential of the medium is very large, the new channel from $gg\gamma$ vertex will assume significance in comparison to the leading order rates.

\begin{figure}[htb]
\begin{center}\vspace{0.0cm}
\includegraphics[width=10cm]{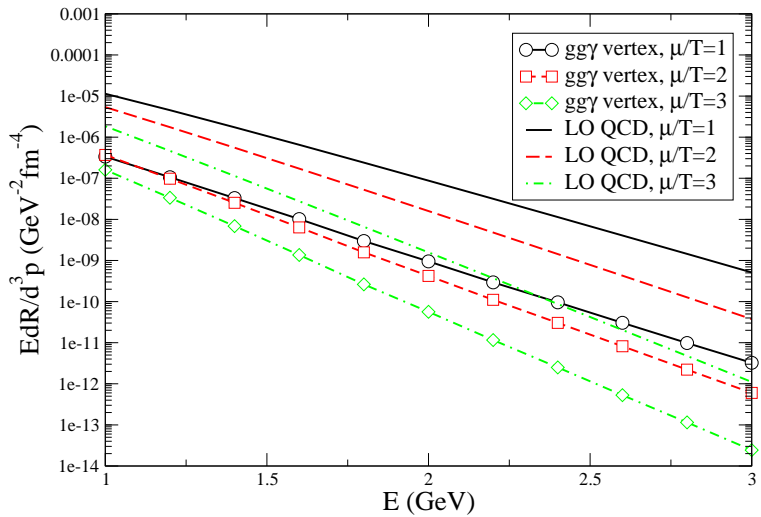}\vspace{0cm}
\end{center}
\caption{The differential rate of photons from $gg\gamma$ vertex in a hot and dense medium with energy density fixed, $\epsilon=1.8GeV/fm^3$, compared with the contribution from QCD annihilation and Compton processes.} \label{epsilon=1.8}
\end{figure}

In the above estimates, the chemical potential and temperature are held fixed separately. In Ref.~\cite{Dumitru:1993us,Vija:1994is}, it is shown that the photon production rate from QCD annihilation and Compton processes have a strong dependence on increasing chemical potential $\mu$ of the medium if the energy density of the medium is fixed. If the energy density were held constant, the temperature $T$ and $\mu$ are related to each other by the equation of state (EOS). In what follows, an estimate of hard photon production from  a medium with fixed energy density is presented and compared with the leading order rates. The equation of state is  derived from the phenomenological MIT bag model~\cite{Chodos:1974je,Dumitru:1993us,Vija:1994is}, where the energy density is given as,
\begin{eqnarray}
\epsilon = A T^4 + CT^2\mu^2 + D\mu^4 + B.
\end{eqnarray}
In the above equation, the constants are given as:
\begin{eqnarray}
A = {37\pi^2 / 30} - {11\pi\alpha_s / 3} \ , \ C = 3(1-2\alpha_s / \pi) \ , \ D = C/(2\pi^2),
\end{eqnarray}
and the bag constant $B$ is fixed to be $200$~MeV$^4$. If $T$ is made dependent of $\mu$ is this way, then both rates will decrease strongly with increasing chemical potential because of the decreasing of the temperature $T$ of the medium. Such a drop is even more pronounced for the case of the leading order rates. At RHIC, one expects a maximum energy density of about $\epsilon=5$~GeV/fm$^3$, and the average energy density will be smaller than this value. We pick a conservative estimate of $\epsilon=1.8$~GeV/fm$^3$, which corresponding to $T=200$~MeV at zero chemical potential. The results of the leading order rate and that from the two gluon channel is presented in Fig.~\ref{epsilon=1.8}.
In the plot, we raise the chemical potential $\mu$ from $\mu/T=1$ to $\mu/T=3$. As is clearly evident, the photon production rate from our new channel decreases with increasing chemical potential with fixed energy density, showing a similar dependence to the QCD annihilation and Compton processes. It would appear that with energy density fixed, the photon production from our new channel has a much stronger dependence on the temperature than on the chemical potential and has not exceeded the photon production rate from QCD annihilation and Compton processes in the range of energies explored. It should be recalled that the above statement is true of only the hard photon production rate and may not hold in an extension to soft photons where cut-cut contributions will contribute. This is left as the subject for a future effort.

\section{Discussions and Conclusions}

A determination of the degrees of freedom prevalent in the early dense matter created in a heavy-ion collision remains an outstanding question in the study of excited strongly interacting matter. While lattice susceptibilities allow for
constraints on the flavor sector of the produced matter, phenomenological explorations remain the sole method to determine the structure of  the gluon sector. As a contribution to this on-going effort we have presented the hard photon signature of the gluon sector. While the gluonic sector may not itself consist of quasi-particle gluons, the production of hard photons through the gluon fusion channel outlined in this paper, similar to the production of large mass Drell-Yan pairs, will be sensitive to the gluon structure functions of such matter.

The matrix element for the conversion of a gluon pair into a photon is vanishing in the vacuum and is non-zero solely in the presence of a charge density in the medium. The amplitude for the production of a photon from two gluons in a thermal environment at finite chemical potential was presented in Sec.~II. This was incorporated into the photon self-energy in a thermalized plasma in Sec.~III. Expressions for the photon production rate from the fusion or decay of in-medium gluons in such a scenario were derived subsequently.

Phenomenological estimates of the photon production rate from such equilibrium channels have tended to be suppressed compared to the leading order rates for realistic values of temperature and chemical potential. Such estimates however do not constrain the photon production rate from jet plasma interaction channels~\cite{Fries:2002kt}. Neither do they limit the rates of photons produced form such channels in non-equilibrium scenarios, such as in the early plasma where the gluon populations far exceed those of the quarks. The results of Sec.~III, which have been cast in the form of a kinetic theory, may be easily extended away from the equilibrium scenarios where they have been derived and applied to the above mentioned situations. Yet another application of such rates is to photon production in neutron stars where the chemical potential far exceeds the temperature and many of the conventional channels are Pauli blocked. Estimates of the photon production rate from two gluons in such diverse scenarios will be presented in upcoming efforts.

\section{Acknowledgements}
This work is supported in part by the Natural Sciences and Engineering Research Council of Canada, and by the U.S. Department of Energy under grant No. DE-FG02-05ER41367 and under Contract No. DE-AC03-76SF00098.


\end{document}